\documentclass[twocolumn]{revtex4}
\usepackage[usenames,dvipsnames]{pstricks}
\usepackage{pst-node}
\usepackage{amsmath}
\usepackage{epsfig}
\usepackage{enumerate}
\usepackage{amssymb}
\usepackage{amscd}

\usepackage{amsfonts, amsthm}
\usepackage[colorlinks=false,breaklinks]{hyperref}
\usepackage{url}

\newcommand{\cV}{\mathcal{V}}
\newcommand{\cG}{\mathcal{G}}
\newcommand{\cH}{\mathcal{H}}

\newcommand{\cD}{\mathcal{D}}

\newcommand{\cL}{\mathcal{L}}
\newcommand{\bra}[1]{\mbox{$\langle #1 |$}}
\newcommand{\ket}[1]{\mbox{$\left| #1 \right\rangle$}}

\newcommand{\ketbra}[2]{\mbox{$\ket{#1}\!\bra{#2}$}}

\newcommand{\proj}[1]{\ketbra{#1}{#1}}
\newcommand{\id}{\mathsf{id}}

\newcommand*{\Hphys}{\cH^{\mbox{\tiny{phys}}}}
\newcommand*{\Ren}{\mathbf{R}}

\newtheorem{lemma}{Lemma}
\newtheorem*{remark}{Remark}
\newtheorem{example}{Example}

\newcommand{\eqnref}[1]{\hyperref[#1]{{(\ref*{#1})}}}
\newcommand{\lemref}[1]{\hyperref[#1]{{Lemma~\ref*{#1}}}}
\newcommand{\figref}[1]{\hyperref[#1]{{Fig.~\ref*{#1}}}}
\newcommand{\appref}[1]{\hyperref[#1]{{Appendix~\ref*{#1}}}}
\newcommand{\exampleref}[1]{\hyperref[#1]{{Example~\ref*{#1}}}}

\def\puncture {\psdots[dotstyle=+,dotangle=45,dotscale=1](0,0)}
\DeclareMathOperator{\reduce}{\operatorname{red}}

\begin{document}

\title{Exact entanglement renormalization for string-net models}

\author{Robert K\"onig}
\affiliation{Institute for Quantum Information, California Institute of Technology}
\author{Ben W.~Reichardt}
\affiliation{School of Computer Science and Institute for Quantum Computing, University of Waterloo}
\author{Guifr\'e Vidal}
\affiliation{School of Physical Sciences, University of Queensland, QLD 4072, Australia} 

\date{\today}

\begin{abstract}
	We construct an explicit renormalization group (RG) transformation for Levin and Wen's string-net models on a hexagonal lattice. The transformation leaves invariant the ground-state ``fixed-point" wave function of the string-net condensed phase. Our construction also produces an exact representation of the wave function in terms of the multi-scale entanglement renormalization ansatz (MERA). This sets the stage for efficient numerical simulations of string-net models using MERA algorithms. It also provides an explicit quantum circuit to prepare the string-net ground-state wave function using a quantum computer.
\end{abstract}


\maketitle

The fractional quantum Hall effect provides the first experimental evidence~\cite{FQHE} for the existence of topological phases of quantum matter. It has motivated the study of topological order and its characterization~\cite{TopOrd}, and has spurred considerable theoretical efforts to find condensed matter systems exhibiting the relevant features, that is, topological ground space degeneracy and anyonic excitations. The interest in concrete Hamiltonian models is  manifold: They provide an important testbed for theoretical concepts such as the topological entanglement entropy~\cite{TopEntropy,TopEntropyLevinWen}, and may serve as a guide to the experimental search for evidence of their existence. Moreover, systems supporting anyons with computationally universal braiding are a promising avenue for the realization of a quantum computer~\cite{Kitaev,QC}.

A fruitful approach to the realization of topological phases is the study of model
systems whose degrees of freedom are  geometric objects such as loops or so-called string-nets (labeled trivalent graphs) embedded in a surface~\cite{LoopString,LevWen}. To respect the topology,
one attempts to find Hamiltonians whose ground states
are topologically invariant, i.e.,  assign equal amplitudes to configurations
that can be smoothly deformed into each other. This invariance property is not sufficient to uniquely fix a topological phase, however. 
To constrain the system further,  it is assumed that the ground states are---as representatives of a particular phase---fixed under a renormalization group (RG) flow, and thus scale-invariant.  A corresponding Hamiltonian consisting of local terms can then be constructed by expressing topological invariance and the ``fixed-point'' property in terms of local constraints~\footnote{That is, linear dependencies between amplitudes of locally differing configurations.}.   Following this program, Levin and Wen~\cite{LevWen} have constructed an exactly soluble ``fixed-point'' Hamiltonian that realizes, starting from an (essentially arbitrary) modular tensor category, a spin Hamiltonian corresponding to the associated doubled (PT-symmetric) topological phase.

The postulated fix-point property of the ground space under RG is a key ingredient of Levin and Wen's construction  as it motivates the choice of the local constraints. However, the outlined procedure does not  provide an RG transformation; in fact, the mere existence of an RG that fixes the ground states is a priori unclear. Here we construct an explicit RG transformation for $(2+1)$-dimensional string-net models with this property. This establishes that the ``fixed-point" wave functions and Hamiltonians of Ref.~\cite{LevWen} are indeed the infrared limit of string-net condensed phases, and thus confirms the validity of the heuristic reasoning underlying~\cite{LevWen}.

The proposed RG transformation can be seen as an instance of~\emph{entanglement renormalization}~\cite{ER}, that is, it proceeds by locally eliminating part of the ground-state entanglement before each coarse-graining step~\footnote{This is essential to avoid increasing the number of degrees of freedom. Our RG transformation  fits nicely into the general theoretical framework of~\cite{ER}, but its derivation is independent of~\cite{ER} and motivated by topological properties of string nets.}. Due to its structure based on local transformation rules, our RG~transformation conserves topological degrees of freedom. In fact, it maps the ground space {\em exactly}  into the ground-space of the coarse-grained system.  These features are analogous to results obtained in~\cite{Aguado} for Kitaev's toric code~\cite{Kitaev} and its generalizations~\footnote{The phases associated with the models studied in~\cite{Aguado} correspond to the modular tensor category associated with the quantum double~$D(G)$ of a finite group~$G$. Note that Levin and Wen's construction does not require this specific form of the category.}. In particular, 
they give rise to an efficient representation of the ground-states as  tensor networks (i.e., in terms of the multi-scale entanglement renormalization ansatz (MERA)~\cite{ER}). They also imply that our RG transformation is a reasonable choice of initial point for numerical (variational) algorithms~\cite{ALGORITHM} when studying, e.g., the stability of topological phases under perturbations. Finally, our RG transformation gives an explicit prescription for efficiently preparing ``fixed-point'' wave functions or reading out topological information using a quantum computer.

\smallskip
Following~\cite{LevWen}, let $\cG$ be a trivalent graph embedded in a surface $S$ so that the components of $S \setminus \cG$ are simply connected (``plaquettes"). The Hilbert space~$\cH_\cG$ of a {\em string-net model} is spanned by the different networks of labeled, oriented strings living on $\cG$'s edges. A standard basis for this space is obtained by orienting $\cG$ and associating to each edge $e$ a Hilbert space $\cV_e \cong \mathbb{C}^{N+1}$ with orthonormal basis $\{\ket{i}_e\}_{i=0}^N$.  Here, $i$ determines the type and direction of string, with $i = 0$ corresponding the absence of a string across edge $e$.  For each $i$, label $i^*$ corresponds to a string of the same type but with the opposite direction; $0^* = 0$.  Then $\cH_\cG = \bigotimes_e \cV_e$. The model is further characterized by {\em branching rules}, the set of triples $\{i,j,k\}$ of string types that are allowed to come together at a vertex, e.g., $\{ i, i^*, 0\}$ is always allowed.  We define the {\em physical subspace} $\Hphys_\cG \subset \cH_\cG$ as the span of all string-net configurations that have an allowed triple at every vertex.

Define a Hamiltonian $H_\cG$ acting on $\cH_\cG$ by 
\begin{equation} \label{eq:HGdef}
H_\cG = - \!\! \sum_{\text{vertices $v$}} \! Q_v \; - \!\! \sum_{\text{plaquettes $p$}} \! B_p \enspace .
\end{equation}
Here, for each vertex $v$, $Q_v$ is the projection onto the set of allowed net edge triples at $v$.  Thus the first term projects onto $\Hphys_\cG$.  The second term has a more complicated definition.  Let $F^{ijm}_{kln}$ be an order-six tensor, indexed by string types, satisfying certain conditions roughly described as self-consistency, unitarity and compatibility with the branching rules; see appendix for full details.  For each plaquette~$p$ the {\em plaquette operator} $B_p$ is a projection on the edges bordering $p$ controlled by the edges with one endpoint on $p$. 
 More precisely, $B_p = \sum_i d_i B_p^i / \sum_i d_i^2$ where $d_i = 1/F^{i i^* 0}_{i i^* 0}$ and $B_p^i$ acts on a simple plaquette $p$ with $r$~boundary edges as 
\begin{equation} \label{eq:bpidefac}
B_p^i \, \Bigg | \raisebox{-1.4em}{\includegraphics{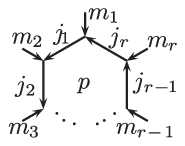}}
\Bigg \rangle
= \!\!
\sum_{k_1,\ldots,k_n} 
\!\!\!\! \bigg( \! \prod_{\nu = 1}^r F^{m_\nu j_\nu^*j_{\nu-1}}_{i^*k_{\nu-1} k_\nu^*} \! \bigg)
\Bigg |\raisebox{-1.4em}{\includegraphics{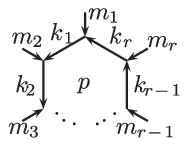}}\Bigg \rangle
\end{equation}
identifying $j_0 = j_r$ and $k_0 = k_r$.  The plaquette and vertex operators commute, and thus the ground space of $H_\cG$ is the space simultaneously fixed by all these projections. 

In the appendix, we give a natural definition of $B_p^i$ for more general plaquettes; roughly,~$B_p^i$ adds a loop of type $i$ around a puncture in the center of $p$ followed by reduction to the basis of $\cH_\cG$.  Eq.~\eqnref{eq:bpidefac} is a special case.  

We now focus on the case where $\cG$ is the honeycomb lattice $\cL$.  Our RG transformation is a map $\Ren: \cH_\cL \rightarrow \cH_{\tilde\cL}$, where $\tilde\cL$ is a coarser hexagonal lattice, that satisfies: 
\begin{enumerate}[(i)]
\item The physical subspace $\cH^{\mbox{\tiny{phys}}}_\cL$ is mapped into $\cH^{\mbox{\tiny{phys}}}_{\tilde\cL}$. \label{it:physicalhspace}
\item Local operators on $\cH_\cL$ are mapped under conjugation by $\Ren$ to local operators on $\cH_{\tilde\cL}$. \label{it:localoperators}
\end{enumerate}
Each plaquette $p$ of $\cL$ is either retained or eliminated by renormalization.  We can show that the form of the plaquette part of the Hamiltonian is preserved under the map~$\Ren$, in the following sense:
\begin{enumerate}[(i)]\setcounter{enumi}{2}
\item\label{it:plaquettetransf}
If $q$ is a retained plaquette of $\cL$ and $\tilde q$ the corresponding plaquette of $\tilde\cL$, then $B_q \big|_{\cH^0_\cL} = \Ren^\dagger B_{\tilde q} \Ren \big|_{\cH^0_\cL}$, where $\cH^0_\cL \subset \Hphys_\cL$ is the subspace simultaneously fixed by all $B_p$ operators for eliminated plaquettes~$p$. 
\end{enumerate}
Furthermore, 
\begin{enumerate}[(i)]\setcounter{enumi}{3}
\item The ground space of $H_\cL$ is mapped bijectively to the ground space of $H_{\tilde\cL}$. 
\label{it:groundstateren}
\end{enumerate}

\begin{figure}
\includegraphics[scale=0.8]{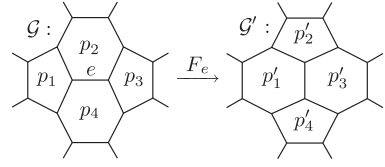}
	\caption{An $F$-move reconnecting an edge $e$ of $\cG$.  Plaquettes of $\cG$ and of $\cG'$ are in one-to-one correspondence.} \label{fig:Fmove}
\end{figure}

The map $\Ren$ is defined by a sequence of \emph{$F$-moves}, elementary trivalent graph transformations.  As shown in \figref{fig:Fmove}, $F_e(\cG)$ is a graph $\cG'$ that is the same as $\cG$ except with an edge $e$ reconnected in a way that corresponds to flipping an edge in the dual graph.  Using the tensor~$F^{ijm}_{kln}$, $F_e$ also defines a linear transformation $\cH_\cG \rightarrow \cH_{\cG'}$, controlled by the labels $\ket{ijkl}$ of the edges adjacent to $e$:  
\begin{equation} \label{eq:Fmovedef}
F_e \, 
\ket{\raisebox{-1.0em}{\includegraphics{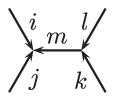}}}
= 
\sum_n F^{ijm}_{kln} \, 
\ket{\raisebox{-1.0em}{\includegraphics{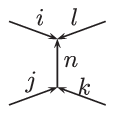}}}
\end{equation}
in the standard string-net bases defined above.  For each edge $e$, $F_e$ maps $\Hphys_\cG$ isomorphically to $\Hphys_{\cG'}$, and $F_e \big|_{\Hphys_\cG}$ can be extended to a unitary on $\cH_\cG$.   

\begin{figure}
\includegraphics[scale=0.8]{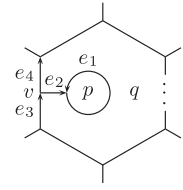}
	\caption{When $\cG$ contains a tadpole around plaquette~$p$ (attached to vertex~$v$) and the state is in the range of $B_p$, it is a product state with respect to the bipartition $\cG\backslash\{e_1,e_2\}:\{e_1,e_2\}$ (\lemref{t:tadpoleremoval}).  In this diagram, the $e_i$ are names for the directed edges and not string-net labels.}
	\label{fig:tadpole}
\end{figure}

\begin{figure*}
	\raisebox{-1.33cm}{\includegraphics[scale=.17]{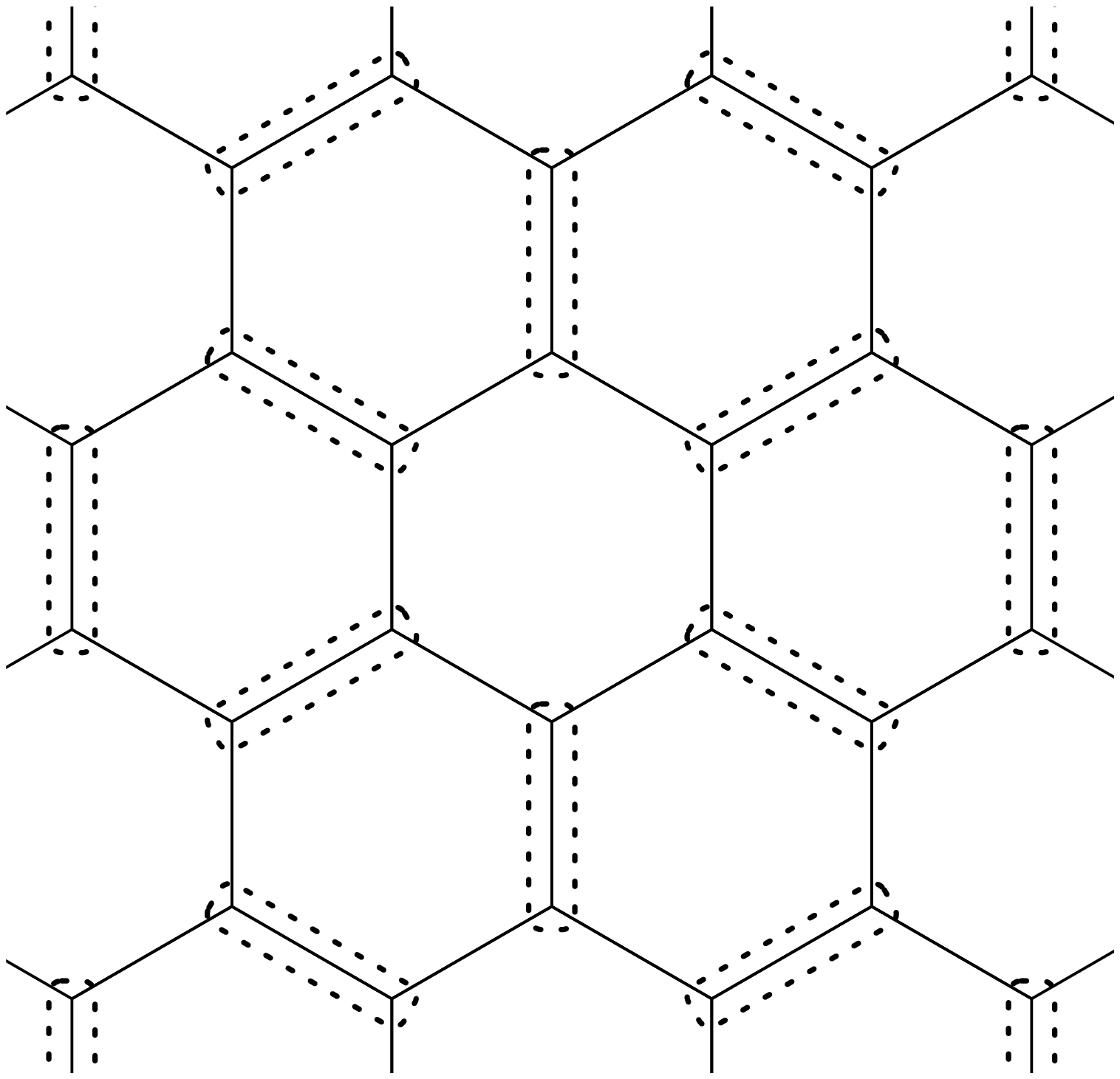}} $\overset{F}{\rightarrow}$
	\raisebox{-1.33cm}{\includegraphics[scale=.17]{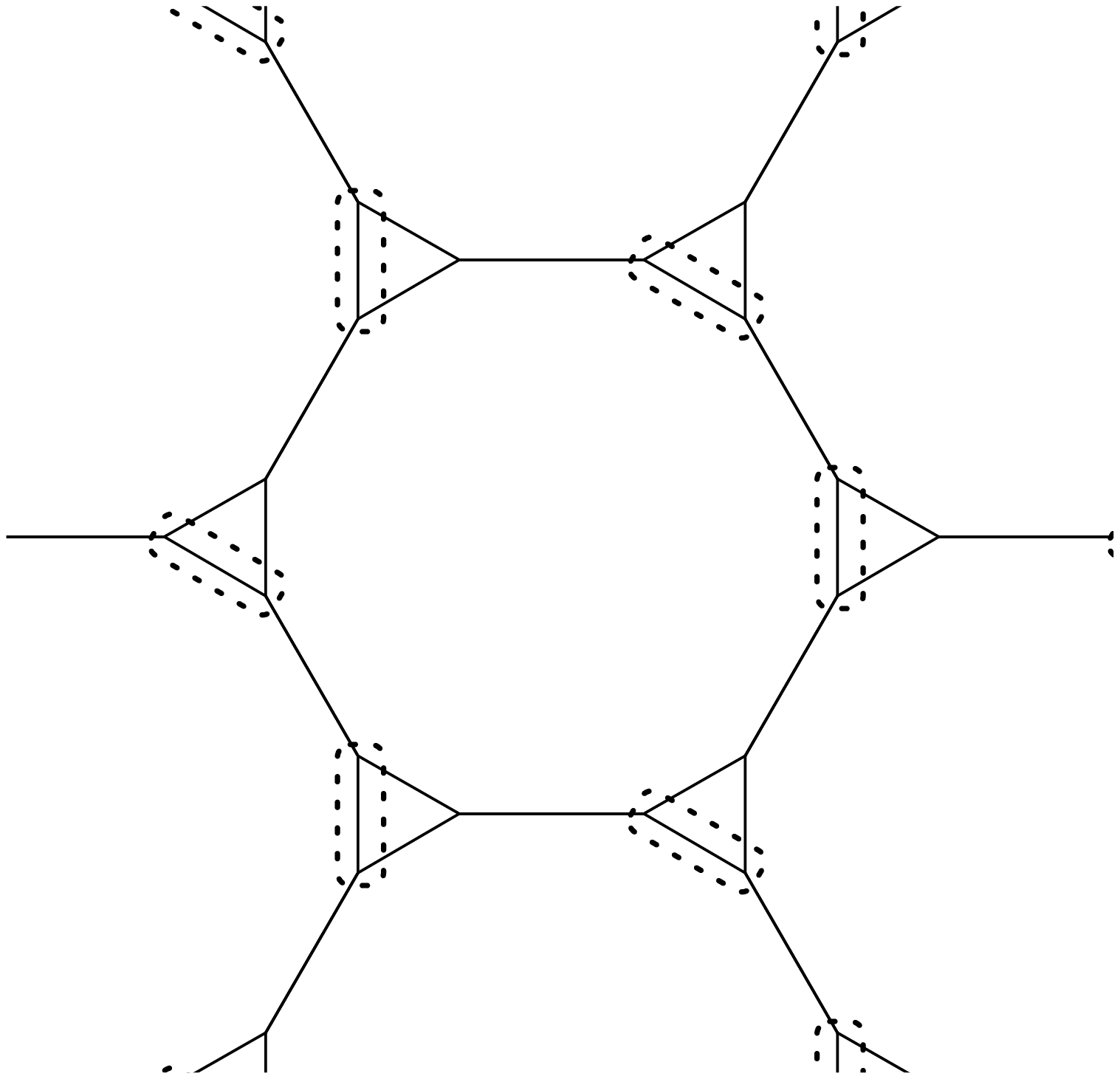}} $\overset{F}{\rightarrow}$
	\raisebox{-1.33cm}{\includegraphics[scale=.17]{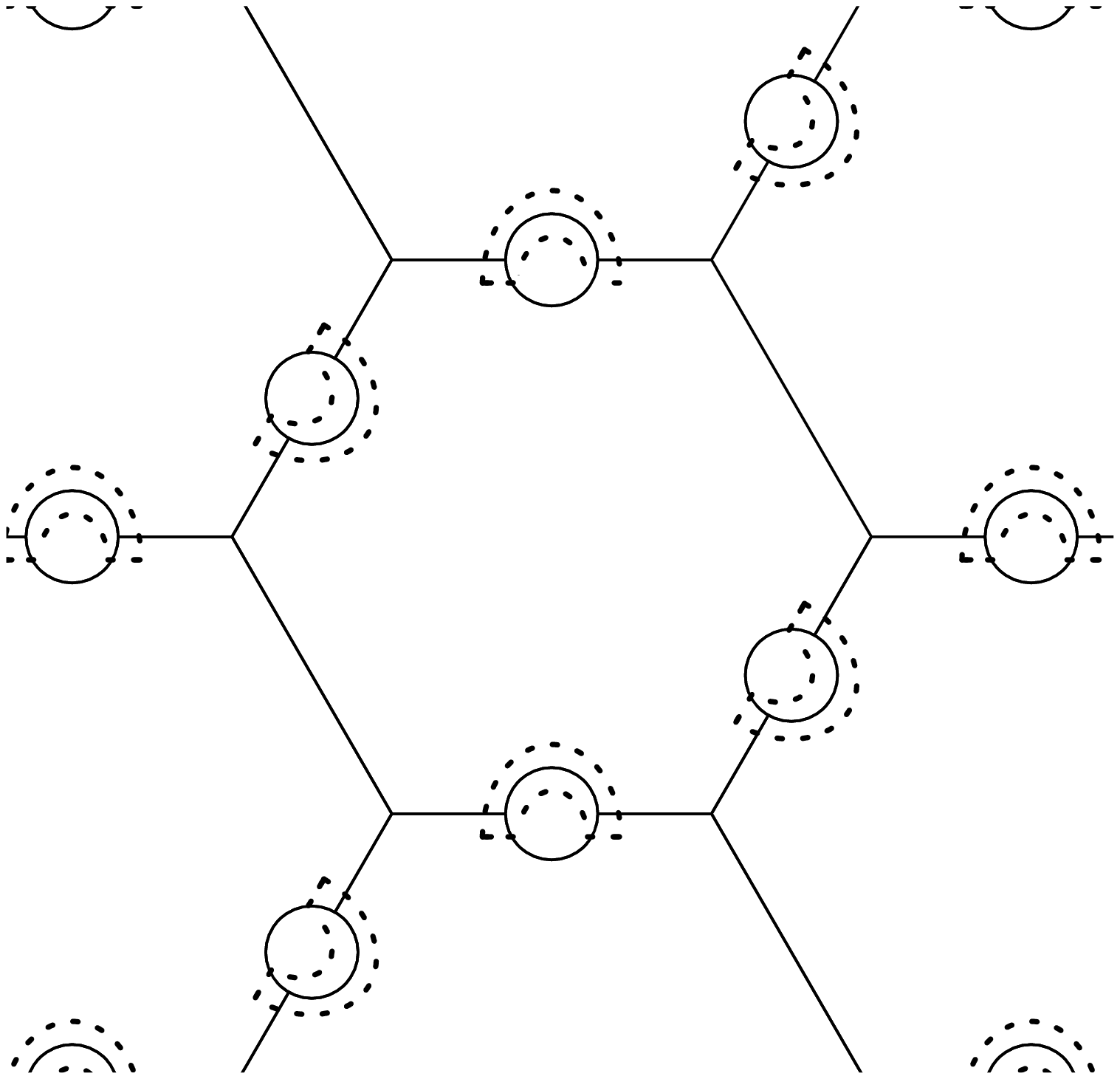}} $\overset{F}{\rightarrow}$
	\raisebox{-1.33cm}{\includegraphics[scale=.17]{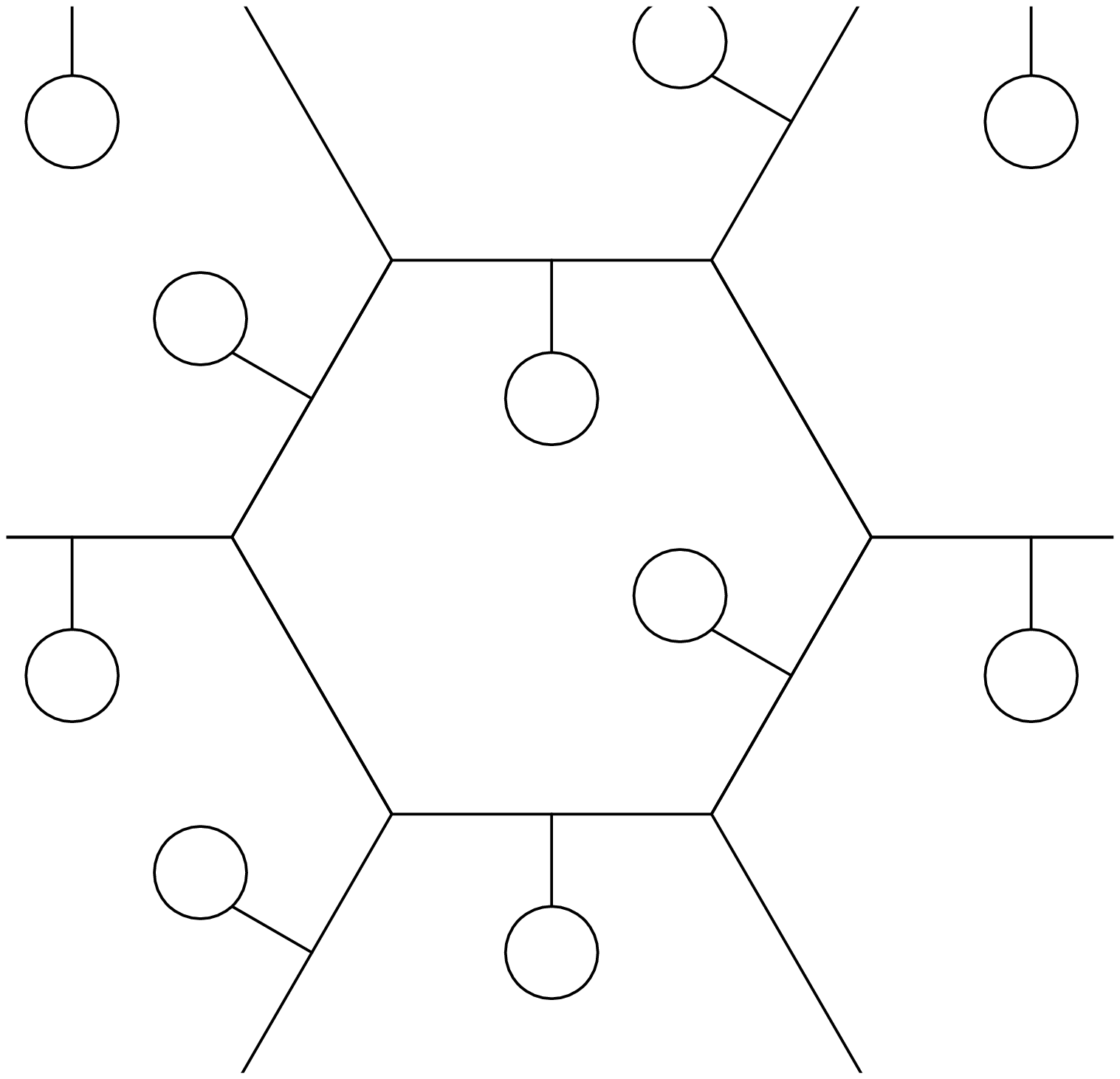}} $\overset{Z}{\rightarrow}$
	\raisebox{-1.33cm}{\includegraphics[scale=.16]{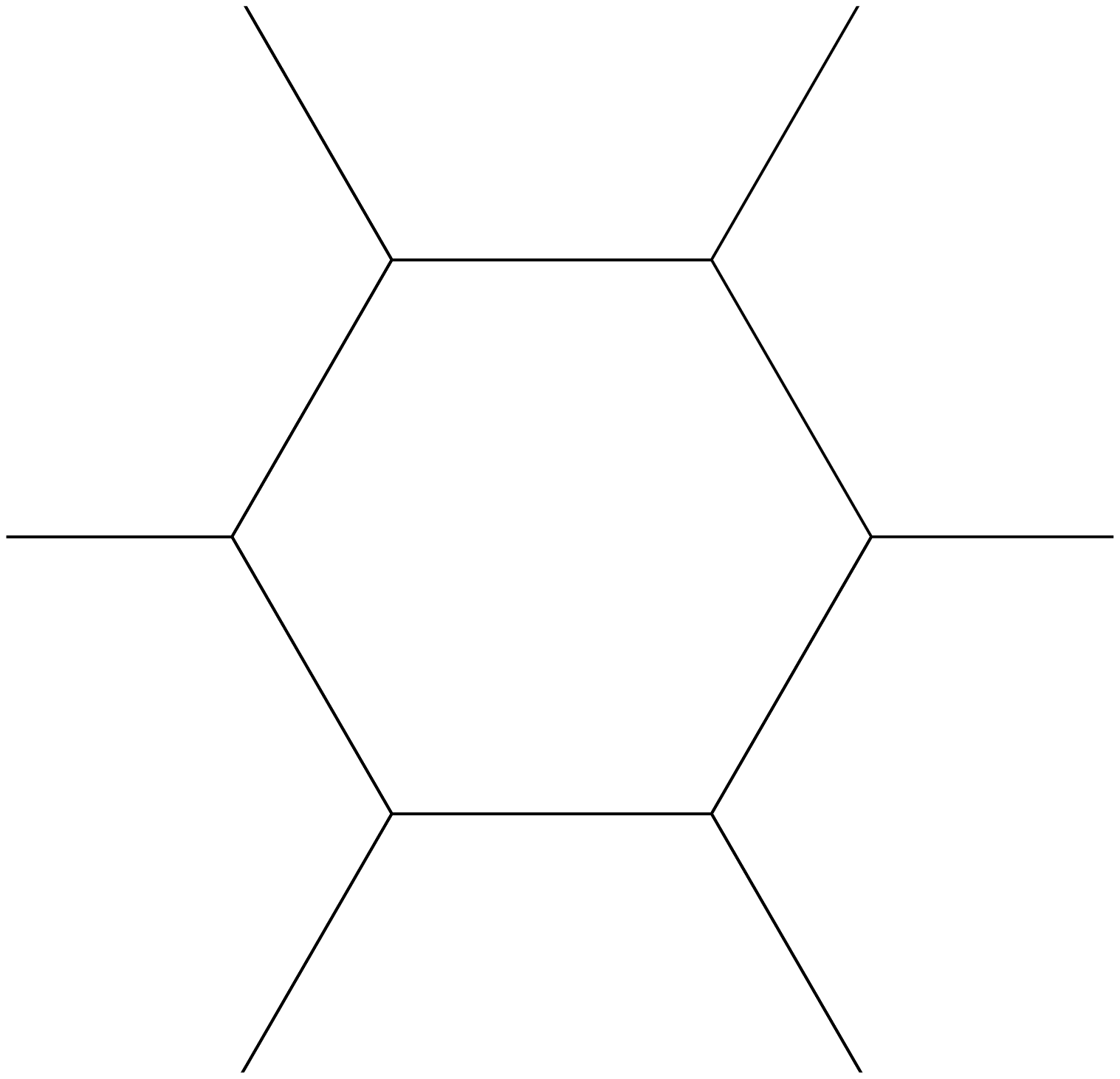}}
	\caption{The RG transformation $\Ren$ coarse-grains lattice $\cL$ into $\tilde\cL$. Edges where $F$-moves are applied are marked by dots. Note that there are many alternative sequences of moves that work equally well.} \label{fig:RGTransformation}
\end{figure*}

A second ingredient of~$\Ren$ are transformations that reduce the number of degrees of freedom by eliminating edges. Suppose that after some $F$-moves, the resulting graph $\cG$ contains a ``tadpole," i.e., a subgraph of the form shown in \figref{fig:tadpole}, consisting of a self-loop around plaquette~$p$, and three other edges. We associate with this tadpole the local operator $Z_p: \cH_\cG \rightarrow \cH_{\cG'}$, where $\cG'$ is obtained from $\cG$ by deleting the tadpole subgraph and replacing edges $e_3$, $e_4$ by a single edge $e'$: 
\begin{equation} \label{eq:Zp}
	Z_p = \bra{\Phi}_{e_1} \otimes \bra{0}_{e_2} \otimes \sum_{i} \ket{i}_{e'} \bra{ii}_{e_3e_4} \otimes \id_{\cG\backslash\{e_1,\ldots,e_4\}}  \ ,
\end{equation}
where $\ket\Phi = \frac1{\sqrt{\sum_i d_i^2}} \sum_i d_i \ket i$.  
Observe $Z_p^\dagger$ is an isometry.  

The map~$\Ren$ from the lattice $\cL$ into the coarser lattice $\tilde\cL$ is now given by the sequence of $F$-moves indicated in \figref{fig:RGTransformation}, followed by eliminating the tadpoles using the $Z_p$ maps. 

The properties of the map~$\Ren$ rely on two basic claims about the behavior of plaquette operators under $F$-moves and the removal of tadpoles. We show that 

\begin{lemma} \label{t:Fcommute}
	For every edge~$e$ and plaquette~$p$, 
	\begin{equation} \label{eq:bpfmove}
 		F_e B_{p} = B_{p'} F_e \enspace ,
	\end{equation}
	where $p'$ corresponds to the plaquette $p$ in the graph $\cG' = F_e(\cG)$.  Roughly speaking, $F$-moves ``commute" with plaquette operators.
\end{lemma}

\lemref{t:Fcommute} implies that the plaquette part $H_\cG$ is mapped to the plaquette part of $H_{\cG'}$ under conjugation by~$F_e$. A similar statement applies to the removal of a tadpole with head $p$ inside a plaquette~$q$; this operation  ``commutes'' with $B_q$ provided we restrict to the subspace fixed by~$B_p$.

\begin{lemma} \label{t:tadpoleremoval}
	Consider a tadpole around plaquette $p$ inside a plaquette~$q$ as shown in~\figref{fig:tadpole}, and let $q'$ be the modified plaquette after removal of the tadpole. Then $B_p$ is a rank-one projection, 
	\begin{equation} \label{eq:bpproj}
		B_p = \proj{\Phi}_{e_1} \otimes \proj{0}_{e_2} \otimes \id_{\cG\backslash\{e_1,e_2\}}  
	\end{equation}
	with $\ket\Phi$ defined as in Eq.~\eqnref{eq:Zp}, and 
	\begin{equation} \label{eq:bptadpolecommutation}
		B_q Q_v B_p 
		= Z_p^\dagger B_{q'} Z_p
		\enspace .
	\end{equation}
	Every ground state $\ket{\Psi}_\cG$ of $H_\cG$ is a product state,\!\!
		\begin{equation} \label{eq:lemma3c}
		\begin{split}
			\ket{\Psi}_\cG 
			&= Z_p^\dagger \ket{\Psi'}_{\cG'} \\
			&= \ket{\Phi}_{e_1} \otimes \ket{0}_{e_2} \otimes \bigg( \sum_i \ket{ii}_{e_3e_4} \! \bra{i}_{e'} \bigg) \ket{\Psi'}_{\cG'}
		\end{split}
		\end{equation}
	where $\ket{\Psi'}_{\cG'}$ is a ground state of $H_{\cG'}$.
\end{lemma}

Lemmas~\ref{t:Fcommute} and~\ref{t:tadpoleremoval} can in principle be verified directly from the explicit expression~\eqnref{eq:bpidefac} for the plaquette operators in terms of standard basis vectors.  A simpler proof is based on the interpretation of $B_p^i$ as adding a ``virtual loop'' to the surface as explained in~\cite[Appendix~C]{LevWen}. The consistency of this interpretation is guaranteed by {\em Mac Lane's coherence theorem}~\cite{MacLane98}, which shows the required reductions yield the same result independently of the sequence of local rules applied. In terms of this interpretation, \lemref{t:Fcommute}  is immediate since the virtual loops are added in a region that is not affected by $F$-moves. Similarly, \lemref{t:tadpoleremoval} follows since the operator $B_p$ effectively removes a puncture in the surface located at the center of~$p$. We present these details and the proofs in the appendix.  

Let us now justify properties~\eqnref{it:physicalhspace}-\eqnref{it:groundstateren} of $\Ren$. 
It is easy to check that both $F$-moves as well as the operators~$Z_p$ preserve the branching rule at every vertex; this proves~\eqnref{it:physicalhspace}.  Similarly,~\eqnref{it:localoperators} immediately follows from the fact that~$\Ren$ is made of local operations. Statement~\eqnref{it:plaquettetransf} is a direct consequence of Lemmas~\ref{t:Fcommute} and~\ref{t:tadpoleremoval}, since Eq.~\eqnref{eq:bptadpolecommutation} implies $B_q \big|_{\cH^0_\cG} = Z_p^\dagger B_{q'} Z_p \big|_{\cH^0_{\cG'}}$. For property~\eqnref{it:groundstateren}, note that the three rounds of $F$-moves in $\Ren$ are unitaries. Therefore we only need to check that $Z_p$, removing a tadpole around $p$ from a graph $\cG$, is a bijection from the ground space of $H_\cG$ to the ground space of $H_{\cG'}$. Again, this directly follows from~\lemref{t:tadpoleremoval}~\footnote{Indeed, if $\ket{\Psi}_\cG$ is a ground state of $H_\cG$, then by \lemref{t:tadpoleremoval}, $\ket{\Psi}_\cG = Z_p^\dagger \ket{\Psi'}_{\cG'}$, and thus $Z_p \ket\Psi_\cG = \ket{\Psi'}_{\cG'}$, for some ground state $\ket{\Psi'}_{\cG'}$ of $H_{\cG'}$. Conversely, if $\ket{\Psi'}_{\cG'}$ is a ground state of $H_{\cG'}$, then let $\ket{\Psi}_\cG = Z_p^\dagger \ket{\Psi'}_{\cG'}$ so $\ket{\Psi'}_\cG = Z_p \ket{\Psi}_\cG$. Eq.~\eqnref{eq:bptadpolecommutation} implies that $Z_p^\dagger B_{q'} = B_q Q_v B_p Z_p^\dagger$, so $\ket{\Psi}_\cG = B_q Q_v B_p \ket{\Psi}_\cG$. Thus $\ket{\Psi}_\cG$ is fixed by all plaquette and vertex operators, so is a ground state of $H_\cG$. Alternatively, we could have used the fact that~$H_\cG$ and $H_{\cG'}$ have the same ground-space degeneracy.}.

Let us remark that Lemmas~\ref{t:Fcommute} and~\ref{t:tadpoleremoval} generalize considerably.  In particular, Property~\eqref{it:plaquettetransf} holds even if~$B_q$ is replaced by the more general Wilson loop operators discussed in~\cite{LevWen} that can act nontrivially on the ground space.  The operator $Z_p^\dagger$ is a special case of \emph{surgery} between two surfaces, one of which is the sphere in this case.  A version of \lemref{t:tadpoleremoval} holds for general surgery.  

\smallskip 

Every iteration of the RG transformation $\Ren$ reduces the number of sites of the lattice $\cL$ by one-third. In the case that $\cL$  is embedded in the infinite plane, the unique ground state $\ket{\Psi}_\cL$ is a fixed point of~$\Ren$ (by property~\eqref{it:groundstateren}).  More interesting are cases with a topological ground space degeneracy, e.g., a finite system on a torus~\footnote{The degeneracy is a function of the genus and the fusion rules of the tensor category. It can be computed by considering, as in~\cite{TopEntropyLevinWen}, a minimal set of inequivalent string-net configurations.  For example, for the Fibonacci model, which has one nontrivial string label, the ground space on the torus is four dimensional. This corresponds to the different configurations of strings along the two fundamental $1$-cycles.}. A ground state $\ket{\Psi}_\cL$ of $H_\cL$ is
eventually reduced to a ground state $\ket{\Psi}_{top}$ of an effective Hamiltonian on a small number of edges; both the state and the Hamiltonian encode the topological features of the original state/model.

 In the terminology of entanglement renormalization~\cite{ER}, we can think of $\Ren$ as being made of \emph{disentanglers} $U: \cV^{\otimes 5} \rightarrow \cV^{\otimes 5}$ (e.g., the first round of $F$-moves) and \emph{isometries} $W: \cV^{\otimes 6} \rightarrow \cV^{\otimes 3}$ (the remaining $F$- and $Z$-moves). $W$ replaces a triangle by a single vertex. This pattern of operations has also been applied in the context of an RG transformation for classical partition functions~\cite{TRG}. By reversing $\Ren$, we obtain an explicit, logarithmic-depth quantum circuit $\mathcal{C}$ to prepare $\ket{\Psi}_\cL$ from $\ket{\Psi}_{top}$ using local gates~\footnote{Observe that the only required ingredients for this circuit are local gates performing $F$-moves and the preparation of single-particle states $\ket{0}_e$ in the vacuum (cf.~\lemref{t:tadpoleremoval}). The circuit $\mathcal{C}$ is an exact MERA~\cite{ER} of the ground-state wave function $\ket{\Psi}_\cL$ of the string-net condensed phase.}.  
This  is a consequence of the recursive character of the RG transformation. It  should be contrasted with~\cite{Bravyietal}, where it is shown that the creation of a topologically ordered state takes a time linear in the system size if it is based on local Hamiltonian evolution.

In summary, the RG transformation presented here  provides
 both a theoretical foundation and a concrete tool for the study of  string-net condensation as a model for topologically ordered phases. Its simple
description in terms of the underlying tensor category translates into an efficient representation of the ground-states. This gives a theoretical indication of the suitability of appropriate numerical RG procedures  in the study of topologically ordered systems, thereby adding to the evidence for their remarkable precision~\cite{ALGORITHM}.

\acknowledgements

We thank Miguel Aguado, Lukasz Fidkowski, Alexei Kitaev, Greg Kuperberg and John Preskill for helpful conversations.  R.K. and B.R.~acknowledge support from NSF Grants CCF-0524828, PHY-0456720, PHY-0803371 and ARO Grant W911NF-05-1-0294. G.V.~acknowledges support~from Australian Research Council (FF0668731, DP0878830).

\appendix

\section{Basic definitions for general string-net models} \label{s:definitions}

We first review the properties that the tensor $F^{ijm}_{kln}$ needs to satisfy in order to define a string-net model.  Start by encoding the branching rules into a tensor $\delta_{ijk}$, with $\delta_{ijk} = 1$ if string types $i$, $j$, $k$ are allowed to come together at a vertex, and $\delta_{ijk} = 0$ otherwise.  The branching rules are assumed to satisfy $\delta_{ij^*0} = \delta_{i j}$, where $\delta_{ij}$ is the Kronecker delta.  Assume that the $F$ tensor satisfies for all $i, j, \ldots, s$:
\begin{eqnarray}
	\!\!\!\!\!\!\begin{array}{r}\mbox{\small{physicality:}}\end{array} \!\!\!\!& 
		F^{ijm}_{kln} \delta_{ijm} \delta_{klm^*} = F^{ijm}_{kln} \delta_{iln} \delta_{jkn^*}
		\label{eq:physicality} \\
	\!\!\!\!\!\!\begin{array}{r} \mbox{\small{pentagon\ }}\\ \mbox{\small{identity:}} \end{array} \!\!\!\!&
		\sum_{n=0}^N F^{mlq}_{kpn} F^{jip^*}_{mns} F^{jsn}_{lkr} = F^{jip^*}_{q^*kr} F^{r^*iq^*}_{mls}
		\label{eq:pentagon} \\
	\!\!\!\!\!\!\begin{array}{r}\mbox{\small{unitarity:}}\end{array} \!\!\!\!& 
		(F^{ijm}_{kln})^* = F^{i^*j^*m^*}_{k^*l^*n^*}
		\label{eq:unitarity} \\
	\!\!\!\!\!\!\begin{array}{r}\mbox{\small{tetrahedral\ }}\\ \mbox{\small{symmetry:}} \end{array} \!\!\!\!&
		F^{ijm}_{kln} = F^{jim}_{lkn^*} = F^{lkm^*}_{jin}=F^{imj}_{k^*nl} \sqrt{ \frac{d_m d_n}{d_j d_l} }
		\label{eq:tetrahedron}\\
	\!\!\!\!\!\!\mbox{\small{normalization:}}\!\!\!\! \!\!\!\!&
		F^{ii^*0}_{j^*jk} = \sqrt{\frac{d_k}{d_i d_j}} \delta_{ijk} 
		\label{eq:normalization}
\end{eqnarray}
where $d_i^{-1} = F^{ii^*0}_{ii^*0} \neq 0$.  
Then via Eqs.~\eqnref{eq:HGdef},~\eqnref{eq:bpidefac} and 
\def\starv {
\scalebox{.7}{ 
\raisebox{0em}{
\begin{pspicture}[shift=-.4](-.606,-.433)(.606,.7) 
\psset{unit=.7cm} 
\psset{linewidth=.8pt} 
\psset{labelsep=2.5pt} 
\SpecialCoor
\pnode(0,0){O}
\psline{<-}(0,0)(1;90)
\psline{<-}(0,0)(1;-30)
\psline{<-}(0,0)(1;210)
\uput{3.5pt}[-90](O){$v$}
\uput[0](.9;90){$i$}
\uput[120](.8;-150){$j$}
\uput[60](.8;-30){$k$}
\end{pspicture}
}}}
\begin{align}
Q_v &= \sum_{i,j,k} \delta_{ijk} \Big| \!\!\starv \Big\rangle \! \Big\langle \!\!\starv \Big| \label{eq:qvdef} \enspace ,
\end{align}
the tensor $F^{ijm}_{kln}$ gives rise to a Hamiltonian~$H_\cL$ of a string-net model on the honeycomb lattice~$\cL$~\cite{LevWen}. To define $H_\cG$ for more general trivalent graphs~$\cG$, though, we need to extend the definition~\eqnref{eq:bpidefac} of the operators~$B_p^i$ to arbitrary plaquettes.

Recall that $\cG$ is embedded in a surface $S$.  Put a puncture in the interior of each plaquette of $\cG$, and let $S^*$ be the resulting punctured surface.  A {\em smooth string net} is an equivalence class of directed trivalent graphs embedded in~$S^*$, where the edges carry string labels (cf.~\cite[Appendix~C]{LevWen} for the case of the honeycomb lattice).  The equivalences consist of isotopy, i.e., smooth deformations of the embedding in $S^*$ (for example, crossing punctures is not allowed), and of reversing the direction of an edge labeled $i$ while changing the label to $i^*$.  

Any smooth string net representative embedded in $\cG \subset S^*$ can be associated with one of the basis vectors of $\cH_\cG = \bigotimes_e \cV_e$ in the natural way, assigning $\ket 0$ for any edge not crossed by the smooth string net. More generally, every smooth string net on $S^*$ uniquely determines an element of~$\cH_\cG$ by applying some sequence of the following local substitution rules to obtain a linear combination of smooth string nets in $\cG$: 

\newcommand*{\tadpolerule}{
\begin{pspicture}[shift=-.425](-1.08,-.43)(1.08,.43)
\psset{unit=.6cm} 
\psset{linewidth=.56pt} 
\psset{labelsep=2.5pt} 
\SpecialCoor
\psline{->}(-1.8660254037844386,0)(-.8660254037844386,0)
\psarc{<-}(0,-.5){1}{30}{150}
\psarc{->}(0,.5){1}{-150}{-30}
\psline{->}(.8660254037844386,0)(1.8660254037844386,0)
\uput[90](-1.4,0){$i$}
\uput[90](1.4,0){$j$}
\uput[90](0,.5){$k$}
\uput[90](0,-.5){$l$}
\end{pspicture}}
\begin{align}
\begin{pspicture}[shift=-.4](-1.25,-.5)(1.25,.5)
\psset{unit=.7cm}
\psset{linewidth=.56pt} 
\psset{labelsep=2.5pt} 
\SpecialCoor
\pnode(-.5,0){A}
\pnode(.5,0){B}
\pnode([nodesep=1,angle=120]A){Aa}
\pnode([nodesep=1,angle=-120]A){Ab}
\pnode([nodesep=1,angle=60]B){Ba}
\pnode([nodesep=1,angle=-60]B){Bb}
\pscurve{->}(Aa)(A)(Ab)
\pscurve{->}(Ba)(B)(Bb)
\rput(A){\uput[30](.5;110){$i$}}
\rput(B){\uput[150](.5;60){$j$}}
\end{pspicture}
&= 
\begin{pspicture}[shift=-.4](-1.25,-.5)(1.25,.5)
\psset{unit=.7cm}
\psset{linewidth=.56pt} 
\psset{labelsep=2.5pt} 
\SpecialCoor
\pnode(-.5,0){A}
\pnode(.5,0){B}
\psline{-}(A)(B)
\psline{->}([nodesep=1,angle=120]A)(A)
\psline{<-}([nodesep=1,angle=-120]A)(A)
\psline{->}([nodesep=1,angle=60]B)(B)
\psline{<-}([nodesep=1,angle=-60]B)(B)
\uput[90](0,0){$0$}
\rput(A){\uput[30](.5;120){$i$}}
\rput(A){\uput[-30](.5;-120){$i$}}
\rput(B){\uput[-150](.5;-60){$j$}}
\rput(B){\uput[150](.5;60){$j$}}
\end{pspicture} \\
\raisebox{-1.0em}{\scalebox{1.0}{
\begin{pspicture}(-.4,-.4)(.4,.4)
\psset{linewidth=.56pt} 
\SpecialCoor
\psarc{->}(0,0){0.35}{0.0}{0.0}
\psset{labelsep=2.5pt} 
\uput[150](.35;150){$i$}
\end{pspicture}
}}
&= d_i \label{e:looprule} \\
\tadpolerule &= \delta_{ij} \, \tadpolerule \label{e:tadpolerule} 
\end{align}\begin{align}
\begin{pspicture}[shift=-.5](-1.25,-.5)(1.0,.5)
\psset{unit=.7cm}
\psset{linewidth=.56pt} 
\psset{labelsep=2.5pt} 
\SpecialCoor
\pnode(-.5,0){A}
\pnode(.5,0){B}
\psline{<-}(A)(B)
\psline{->}([nodesep=1,angle=120]A)(A)
\psline{->}([nodesep=1,angle=-120]A)(A)
\psline{->}([nodesep=1,angle=60]B)(B)
\psline{->}([nodesep=1,angle=-60]B)(B)
\uput[90](0,0){$m$}
\rput(A){\uput[30](.5;120){$i$}}
\rput(A){\uput[-30](.5;-120){$j$}}
\rput(B){\uput[-150](.5;-60){$k$}}
\rput(B){\uput[150](.5;60){$l$}}
\end{pspicture}
&= 
\sum_n F^{ijm}_{kln}
\begin{pspicture}[shift=-.75](-.5,-.8)(.5,.8)
\psset{unit=.7cm}
\psset{linewidth=.56pt} 
\psset{labelsep=2.5pt} 
\SpecialCoor
\pnode(0,.5){A}
\pnode(0,-.5){B}
\psline{<-}(A)(B)
\psline{->}([nodesep=1,angle=30]A)(A)
\psline{->}([nodesep=1,angle=150]A)(A)
\psline{->}([nodesep=1,angle=-30]B)(B)
\psline{->}([nodesep=1,angle=-150]B)(B)
\uput[0](0,0){$n$}
\rput(A){\uput[60](.5;150){$i$}}
\rput(B){\uput[120](.5;-150){$j$}}
\rput(B){\uput[60](.5;-30){$k$}}
\rput(A){\uput[120](.5;30){$l$}}
\end{pspicture} \label{e:Frulesmooth}
\end{align}
Crucially, the element of $\cH_\cG$ obtained in this fashion is independent of which sequence of local rules was applied. This self-consistency of the local rules is a special case of Mac Lane's coherence theorem~\cite{MacLane98} (see also~\cite[Appendix~E]{KitaevAnyons}). 

Now define $B_p^i$ as adding a counterclockwise oriented loop with label~$i$ around the puncture in~$p$, followed by reduction back to the standard basis of $\cH_\cG$.  It is straightforward to derive Eq.~\eqnref{eq:bpidefac} from this more general definition (\exampleref{t:bpexample}).  This completes the definition of $H_\cG$ for general trivalent graphs~$\cG$.

\begin{example} \label{t:trivalentbubble}
A smooth string-net ``bubble" with three incoming edges and no interior punctures can be simplified to a trivalent vertex by, e.g., applying an $F$-move to the edge labeled $l$, using Eqs.~\eqnref{e:Frulesmooth} and \eqnref{e:tadpolerule}, followed by applying an $F$-move to the edge labeled $m$ and simplifying with Eqs.~\eqnref{e:tadpolerule}, \eqnref{e:looprule} and~\eqnref{eq:normalization}:
\begin{align} \label{e:trivalentbubble}
\begin{pspicture}[shift=-.8](-1.4,-.8)(1.4,1.1)
\psset{unit=.7cm}
\psset{linewidth=.56pt} 
\psset{labelsep=2.5pt} 
\SpecialCoor
\pnode(0,0){O}
\def\trisize{.57735}
\pnode([nodesep=\trisize,angle=-150]O){a}
\pnode([nodesep=1,angle=-150]a){aa}
\pnode([nodesep=\trisize,angle=-30]O){b}
\pnode([nodesep=1,angle=-30]b){bb}
\pnode([nodesep=\trisize,angle=90]O){c}
\pnode([nodesep=1,angle=90]c){cc}
\ncline{->}{aa}{a} \Aput{$i$}
\ncline{->}{bb}{b} \Bput{$j$}
\ncline{->}{cc}{c} \Aput{$k$}
\ncline{->}{a}{b}  \Bput{$l$}
\ncline{->}{b}{c}  \Bput{$m$}
\ncline{->}{c}{a}  \Bput{$n$}
\end{pspicture}
&= 
F^{nil^*}_{jm^*k^*} \;
\begin{pspicture}[shift=-.75](-.86,-.5)(.86,1.1)
\psset{unit=.7cm}
\psset{linewidth=.56pt} 
\psset{labelsep=2.5pt} 
\SpecialCoor
\pnode(0,0){O}
\pnode([nodesep=1,angle=-150]O){a}
\pnode([nodesep=1,angle=-30]O){b}
\pnode([nodesep=.666,angle=90]O){c}
\pnode([nodesep=.666,angle=90]c){cc}
\pnode([nodesep=.666,angle=90]cc){ccc}
\ncline{->}{a}{O} \Aput{$i$}
\ncline{->}{b}{O} \Aput{$j$}
\ncline{->}{c}{O} \Aput{$k$}
\ncline{->}{ccc}{cc}  \Aput{$k$}
\ncarc[arcangleA=-60,arcangleB=-60,ncurv=1]{->}{c}{cc} \Bput{$m$}
\ncarc[arcangleA=-60,arcangleB=-60,ncurv=1]{->}{cc}{c} \Bput{$n$}
\end{pspicture}
\\&=
\sqrt{\frac{d_m d_n}{d_k}} \delta_{ijk} F^{nil^*}_{jm^*k^*} \;
\begin{pspicture}[shift=-.5](-.86,-.5)(.86,1)
\psset{unit=.7cm}
\psset{linewidth=.56pt} 
\psset{labelsep=2.5pt} 
\SpecialCoor
\pnode(0,0){O}
\pnode([nodesep=1,angle=-150]O){a}
\pnode([nodesep=1,angle=-30]O){b}
\pnode([nodesep=1,angle=90]O){c}
\ncline{->}{a}{O} \Aput{$i$}
\ncline{->}{b}{O} \Aput{$j$}
\ncline{->}{c}{O} \Aput{$k$}
\end{pspicture} \nonumber
\end{align}
\end{example}

\begin{example} \label{t:bpexample}
The operator $B_p^i$ adds a loop of type~$i$, followed by expanding the resulting smooth string net into a sum of standard basis vectors.  For example, 
\begin{align*}
\scalebox{.7}{
\begin{pspicture}[shift=-1](-1.73205,-1)(1.73205,1.4)
\psset{unit=.7cm}
\psset{linewidth=.8pt} 
\psset{labelsep=2.5pt} 
\SpecialCoor
\pnode(0,0){O}
\def\trisize{1} 
\pnode([nodesep=\trisize,angle=-150]O){a}
\pnode([nodesep=1,angle=-150]a){aa}
\pnode([nodesep=\trisize,angle=-30]O){b}
\pnode([nodesep=1,angle=-30]b){bb}
\pnode([nodesep=\trisize,angle=90]O){c}
\pnode([nodesep=1,angle=90]c){cc}
\ncline{->}{aa}{a} \Aput{$m_1$}
\ncline{->}{bb}{b} \Bput{$m_2$}
\ncline{->}{cc}{c} \Aput{$m_3$}
\ncline{->}{a}{b}  \Bput{$j_1$}
\ncline{->}{b}{c}  \Bput{$j_2$}
\ncline{->}{c}{a}  \Bput{$j_3$}
\rput(O){\puncture}
\psarc{->}(O){.35}{0.0}{0.0}
\psset{labelsep=1.5pt} 
\uput[-150](.35;-160){$i$}
\end{pspicture}
}
&=
\sum_{k_1, k_2, k_3}
\prod_{\nu=1}^{3} F^{i^*i0}_{j_\nu j_\nu^* k_\nu}
\;
\scalebox{.7}{
\begin{pspicture}[shift=-1](-1.73205,-1)(1.73205,1.4)
\psset{unit=.7cm}
\psset{linewidth=.8pt} 
\psset{labelsep=2.5pt} 
\SpecialCoor
\pnode(0,0){O}
\def\trisize{1.1547}
\def\smalltrisize{.57735}
\def\tail{1.57735}
\pnode([nodesep=\trisize,angle=-150]O){A}
\pnode([nodesep=\trisize,angle=-30]O){B}
\pnode([nodesep=\trisize,angle=90]O){C}
\pnode([nodesep=\tail,angle=-150]A){a}
\pnode([nodesep=\tail,angle=-30]B){b}
\pnode([nodesep=\tail,angle=90]C){c}
\pnode([nodesep=\smalltrisize,angle=-150]A){Aa}
\pnode([nodesep=\smalltrisize,angle=-30]A){Ab}
\pnode([nodesep=\smalltrisize,angle=90]A){Ac}
\pnode([nodesep=\smalltrisize,angle=-150]B){Ba}
\pnode([nodesep=\smalltrisize,angle=-30]B){Bb}
\pnode([nodesep=\smalltrisize,angle=90]B){Bc}
\pnode([nodesep=\smalltrisize,angle=-150]C){Ca}
\pnode([nodesep=\smalltrisize,angle=-30]C){Cb}
\pnode([nodesep=\smalltrisize,angle=90]C){Cc}
\ncline{->}{a}{Aa} \Aput{$m_1$}
\ncline{->}{b}{Bb} \Bput{$m_2$}
\ncline{->}{c}{Cc} \Aput{$m_3$}
\ncline{->}{Aa}{Ab} \Bput{$j_1$}
\ncline{->}{Ba}{Bb} \Bput{$j_1$}
\ncline{->}{Bb}{Bc} \Bput{$j_2$}
\ncline{->}{Cb}{Cc} \Bput{$j_2$}
\ncline{->}{Cc}{Ca} \Bput{$j_3$}
\ncline{->}{Ac}{Aa} \Bput{$j_3$}
\ncline{->}{Ab}{Ba} \Bput{$k_1$}
\ncline{->}{Bc}{Cb} \Bput{$k_2$}
\ncline{->}{Ca}{Ac} \Bput{$k_3$}
\ncline{->}{Ac}{Ab} \Aput{$i$}
\ncline{->}{Ba}{Bc} \Aput{$i$}
\ncline{->}{Cb}{Ca} \Aput{$i$}
\rput(O){\puncture}
\end{pspicture}
} \\
&= \sum_{k_1,k_2,k_3} \!\!\!\! \bigg( \! \prod_{\nu = 1}^3 F^{m_\nu j_\nu^* j_{\nu-1}}_{i^* k_{\nu-1} k_\nu^*} \! \bigg) \!\!\!\!
\scalebox{.7}{
\begin{pspicture}[shift=-1](-1.73205,-1)(1.73205,2)
\psset{unit=.7cm}
\psset{linewidth=.8pt} 
\psset{labelsep=2.5pt} 
\SpecialCoor
\pnode(0,0){O}
\def\trisize{1} 
\pnode([nodesep=\trisize,angle=-150]O){a}
\pnode([nodesep=1,angle=-150]a){aa}
\pnode([nodesep=\trisize,angle=-30]O){b}
\pnode([nodesep=1,angle=-30]b){bb}
\pnode([nodesep=\trisize,angle=90]O){c}
\pnode([nodesep=1,angle=90]c){cc}
\ncline{->}{aa}{a} \Aput{$m_1$}
\ncline{->}{bb}{b} \Bput{$m_2$}
\ncline{->}{cc}{c} \Aput{$m_3$}
\ncline{->}{a}{b}  \Bput{$k_1$}
\ncline{->}{b}{c}  \Bput{$k_2$}
\ncline{->}{c}{a}  \Bput{$k_3$}
\rput(O){\puncture}
\psset{labelsep=1.5pt} 
\end{pspicture}
}\end{align*}
Here in the first step we have applied three $F$-moves, and in the second step we have applied Eq.~\eqnref{e:trivalentbubble} three times and simplified.  The puncture in plaquette $p$ is marked by~\protect\includegraphics{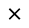}.
Thus we have derived Eq.~\eqnref{eq:bpidefac} for the case that $p$ has $r=3$ sides.  
\end{example}

\begin{remark}
The derivation in \exampleref{t:bpexample} suggests a convenient shorthand rule for determining the action of $B_p^i$.  First, draw a loop with label~$i$ going counterclockwise along the boundary inside plaquette~$p$. Then, formally replace each $T$-junction as shown:
\begin{equation*}
\begin{pspicture}[shift=-2.2em](-1,-.5)(1,1)
\psset{linewidth=.56pt} 
\SpecialCoor
\pnode(0,0){O}
\psline{<-}(O)(1;0)
\psline{<-}(O)(-1;0)
\psline{<-}(O)(1;90)
\psline{->}(1,-.2)(-1,-.2)
\rput([nodesep=.4,angle=-90]O){$i$}
\rput[b]([nodesep=.8,angle=75]O){$a$}
\rput[b]([nodesep=.8,angle=170]O){$b$}
\rput[b]([nodesep=.8,angle=10]O){$c$}
\end{pspicture}
\,\longrightarrow\,
F^{abc}_{i^*c'b'}\;
\begin{pspicture}[shift=-2.2em](-1,-.5)(1,1)
\psset{linewidth=.56pt} 
\SpecialCoor
\pnode(0,0){O}
\psline{<-}(O)(1;0)
\psline{<-}(O)(-1;0)
\psline{<-}(O)(1;90)
\rput[b]([nodesep=.8,angle=75]O){$a$}
\rput[b]([nodesep=.8,angle=170]O){$b'$}
\rput[b]([nodesep=.8,angle=10]O){$c'$}
\end{pspicture}
\end{equation*}
\noindent Finally, identify primed variables at adjacent junctions, and sum over the remaining primed variables.  It is easy to check that this rule computes $B_p^i$, although special care must be taken to apply the rule to a plaquette with degenerate boundary.  
\end{remark}

\section{Proofs of Lemmas~\ref{t:Fcommute} and~\ref{t:tadpoleremoval}} \label{s:proofs}

\begin{proof}[Proof of \lemref{t:Fcommute}]
We claim that $F_e B_p^i = B_{p'}^i F_e$.  Since $B_p^i$ is defined as adding a loop of type $i$ followed by reduction to the standard basis of the graph, this claim is equivalent to the following diagram commuting:
\begin{equation*}
\includegraphics[scale=.85]{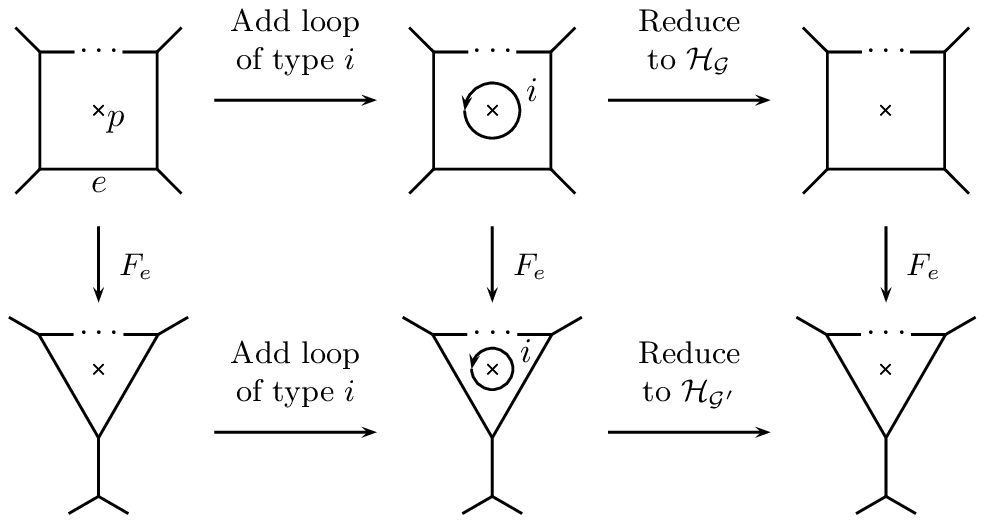} 
\end{equation*}
To simplify the diagram, we have drawn only $\cG$ and $\cG' = F_e(\cG)$, instead of writing superpositions of basis states. 

Now the left half of this diagram commutes since $e$ is separated away from the puncture.  The right half of the diagram commutes by Mac Lane's coherence theorem, since the two ways around it are different ways of reducing to $\cH_{\cG'}$.  
\end{proof}

\noindent Thus \lemref{t:Fcommute} is a nearly immediate corollary of Mac Lane's coherence theorem.  This simple proof shows the usefulness of defining $B_p^i$ using smooth string nets.  A similar argument shows that $[B_p^i, B_q^j] = 0$ for all plaquettes $p$, $q$ and all string-net types $i$, $j$, as we asserted below Eq.~\eqnref{eq:bpidefac}.  

For the proof of \lemref{t:tadpoleremoval}, we first show the following rule that applies to smooth string nets:

\begin{lemma} \label{t:punctureremoval}
\begin{equation}
B_p 
\;
\begin{pspicture}[shift=-.4](-1,-.5)(0,.5)
\psset{linewidth=.56pt} 
\SpecialCoor
\pnode(-.5,0){P}
\pnode(-1,-.5){iA}
\pnode(-1,0){iB}
\pnode(-1,.5){iC}
\pnode(0,0){iBp}
\rput(P){\puncture}
\rput([nodesep=.2,angle=-30]P){$p$}
\pscurve[showpoints=false]{->}(iA)(iB)(iC)
\rput([nodesep=.2,angle=-30]iC){$i$}
\end{pspicture}
=
B_p \;
\begin{pspicture}[shift=-.4](-1,-.5)(-.1,.5)
\psset{linewidth=.56pt} 
\SpecialCoor
\pnode(-.5,0){P}
\pnode(-1,-.5){iA}
\pnode(-1,0){iB}
\pnode(-1,.5){iC}
\pnode(-.1,0){iBp}
\rput(P){\puncture}
\psecurve{->}([nodesep=.25,angle=-90]iA)(iA)(iBp)(iC)([nodesep=.25,angle=90]iC)
\rput([nodesep=.2,angle=30]iC){$i$}
\end{pspicture}
\end{equation}
\end{lemma}

\noindent
Intuitively, \lemref{t:punctureremoval} says that applying $B_p$ effectively removes from $S^*$ the puncture $p$ by allowing strings to be carried over it isotopically.  The proof is by applying two $F$-moves.  Let $\cD = \sqrt{\sum_k d_k^2}$, the ``total quantum dimension."  

\begin{proof}
\newcommand*{\bpdeformlastpart}{
\begin{pspicture}[shift=-.4](-1,-.5)(-.1,.5)
\psset{linewidth=.56pt} 
\SpecialCoor
\pnode(-.5,0){P}
\pnode(-1,-.5){iA}
\pnode(-1,0){iB}
\pnode(-1,.5){iC}
\pnode([nodesep=.25,angle=-90]P){iBa}
\pnode([nodesep=.25,angle=90]P){iBc}
\nccurve[angleA=30,angleB=-120]{->}{iA}{iBa}
\Bput[2.5pt]{$i$}
\nccurve[angleA=120,angleB=-30]{->}{iBc}{iC}
\Bput[2.5pt]{$i$} 
\rput(P){\puncture}
\psarc[arcsepB=2pt]{->}(P){.25}{-90}{90}
\psarc[arcsepB=2pt]{<-}(P){.25}{90}{270}
\rput([nodesep=.4,angle=0]P){$j$}
\rput([nodesep=.4,angle=180]P){$k$}
\psecurve{->}([nodesep=.25,angle=-90]iA)(iA)(iBa)
\psecurve{->}(iBc)(iC)([nodesep=.25,angle=90]iC)
\end{pspicture}}

By definition of $B_p$, 
\begin{align*}
\cD^2 B_p \;
\begin{pspicture}[shift=-.4](-1,-.5)(0,.5)
\psset{linewidth=.56pt} 
\SpecialCoor
\pnode(-.5,0){P}
\pnode(-1,-.5){iA}
\pnode(-1,0){iB}
\pnode(-1,.5){iC}
\pnode(0,0){iBp}
\rput(P){\puncture}
\rput([nodesep=.2,angle=-30]P){$p$}
\pscurve[showpoints=false]{->}(iA)(iB)(iC)
\rput([nodesep=.2,angle=-30]iC){$i$}
\end{pspicture}
&=
\sum_j d_j B_p^j \;
\begin{pspicture}[shift=-.4](-1,-.5)(0,.5)
\psset{linewidth=.56pt} 
\SpecialCoor
\pnode(-.5,0){P}
\pnode(-1,-.5){iA}
\pnode(-1,0){iB}
\pnode(-1,.5){iC}
\pnode(0,0){iBp}
\rput(P){\puncture}
\rput([nodesep=.2,angle=-30]P){$p$}
\pscurve[showpoints=false]{->}(iA)(iB)(iC)
\rput([nodesep=.2,angle=-30]iC){$i$}
\end{pspicture}\\
&=
\sum_j d_j \;
\begin{pspicture}[shift=-.4](-1,-.5)(0,.5)
\psset{linewidth=.56pt} 
\SpecialCoor
\pnode(-.5,0){P}
\pnode(-1,-.5){iA}
\pnode(-1,0){iB}
\pnode(-1,.5){iC}
\pnode(0,0){iBp}
\rput(P){\puncture}
\psarc[arcsepB=2pt]{->}(P){.25}{0}{360}
\rput([nodesep=.45,angle=-30]P){$j$}
\pscurve[showpoints=false]{->}(iA)(iB)(iC)
\rput([nodesep=.2,angle=-30]iC){$i$}
\end{pspicture}
\\&= 
\sum_{j,k} d_j F^{i^* i 0}_{j^* j k} \;
\begin{pspicture}[shift=-.4](-1,-.5)(-.1,.5)
\psset{linewidth=.56pt} 
\SpecialCoor
\pnode(-.5,0){P}
\pnode(-1,-.5){iA}
\pnode(-1,0){iB}
\pnode(-1,.5){iC}
\pnode([nodesep=.25,angle=-90]P){iBa}
\pnode([nodesep=.25,angle=90]P){iBc}
\nccurve[angleA=30,angleB=-120]{->}{iA}{iBa}
\Bput[2.5pt]{$i$}
\nccurve[angleA=120,angleB=-30]{->}{iBc}{iC}
\Bput[2.5pt]{$i$}
\rput(P){\puncture}
\psarc[arcsepB=2pt]{->}(P){.25}{-90}{90}
\psarc[arcsepB=2pt]{<-}(P){.25}{90}{270}
\rput([nodesep=.4,angle=0]P){$j$}
\rput([nodesep=.4,angle=180]P){$k$}
\psecurve{->}([nodesep=.25,angle=-90]iA)(iA)(iBa)
\psecurve{->}(iBc)(iC)([nodesep=.25,angle=90]iC)
\end{pspicture}
\\&=
\sum_{j,k} \sqrt{\frac{d_j d_k}{d_i}} \delta_{i^*jk} \;
\bpdeformlastpart
\\
\intertext{We have made an $F$-move and used Eq.~\eqnref{eq:normalization}.  Every smooth string net depicted above represents the corresponding element of $\cH_\cG$; the use of Mac Lane's theorem is implicit.  Now by symmetry,}
\sum_{j,k} \sqrt{\frac{d_j d_k}{d_i}} \delta_{i^*jk} \;
\bpdeformlastpart
&= 
\sum_k d_k \; 
\begin{pspicture}[shift=-.4](-1,-.5)(-.1,.5)
\psset{linewidth=.56pt} 
\SpecialCoor
\pnode(-.5,0){P}
\pnode(-1,-.5){iA}
\pnode(-1,0){iB}
\pnode(-1,.5){iC}
\pnode(-.2,-.25){iBa}
\pnode(-.1,0){iBb}
\pnode(-.2,.25){iBc}
\rput(P){\puncture}
\psarc[arcsepB=2pt]{<-}(P){.25}{0}{360}
\rput([nodesep=.45,angle=-160]P){$k$}
\psecurve{->}([nodesep=.25,angle=-90]iA)(iA)(iBa)(iBb)(iBc)(iC)([nodesep=.25,angle=90]iC)
\rput([nodesep=.2,angle=0]iBb){$i$}
\end{pspicture}
\\&=
\cD^2 B_p \;
\begin{pspicture}[shift=-.4](-1,-.5)(-.1,.5)
\psset{linewidth=.56pt} 
\SpecialCoor
\pnode(-.5,0){P}
\pnode(-1,-.5){iA}
\pnode(-1,0){iB}
\pnode(-1,.5){iC}
\pnode(-.1,0){iBp}
\rput(P){\puncture}
\psecurve{->}([nodesep=.25,angle=-90]iA)(iA)(iBp)(iC)([nodesep=.25,angle=90]iC)
\rput([nodesep=.2,angle=0]iBp){$i$}
\end{pspicture}
\qedhere
\end{align*}
\end{proof}

\begin{proof}[Proof of \lemref{t:tadpoleremoval}]
First, note that 
\begin{align}
B_p 
\;
\begin{pspicture}[shift=-.4](-1.2,-.4)(.7,.4)
\psset{linewidth=.56pt} 
\SpecialCoor
\pnode(0,0){P}
\pnode([nodesep=.4,angle=180]P){v}
\pnode(-1.2,0){w}
\rput(P){\puncture}
\rput([nodesep=.2,angle=-30]P){$p$}
\psarc[arcsepB=2pt]{->}(P){.4}{180}{-180}
\nccurve[angleA=0,angleB=180]{->}{w}{v}
\Aput[2.5pt]{$j$}
\rput(.6;45){$k$}
\end{pspicture}
&=
B_p 
\;
\begin{pspicture}[shift=-.4](-1.2,-.4)(.4,.4)
\psset{linewidth=.56pt} 
\SpecialCoor
\pnode(0,0){P}
\pnode(-.4,.4){v}
\pnode(-1.2,0){w}
\rput(P){\puncture}
\rput([nodesep=.2,angle=-30]P){$p$}
\psarc[arcsepB=2pt]{->}([nodesep=.25,angle=0]v){.25}{180}{-180}
\nccurve[angleA=0,angleB=180]{->}{w}{v}
\Aput[2.5pt]{$j$}
\rput([nodesep=.25,angle=0]v){\rput(.45;45){$k$}}
\end{pspicture}\nonumber \\
&=
\delta_{j0} \, B_p \;
\begin{pspicture}[shift=-.4](-.4,-.4)(.4,.4)
\psset{linewidth=.56pt} 
\SpecialCoor
\pnode(0,0){P}
\pnode(-.4,.4){v}
\pnode(-1.2,0){w}
\rput(P){\puncture}
\rput([nodesep=.2,angle=-30]P){$p$}
\psarc[arcsepB=2pt]{->}([nodesep=.25,angle=0]v){.25}{180}{-180}
\rput([nodesep=.25,angle=0]v){\rput(.4;0){$k$}}
\end{pspicture}\nonumber \\
&=
\delta_{j0} d_k \, B_p \;
\begin{pspicture}[shift=0](-.1,-.1)(.1,.1)
\psset{linewidth=.56pt} 
\SpecialCoor
\pnode(0,0){P}
\rput(P){\puncture}
\rput([nodesep=.2,angle=-30]P){$p$}
\end{pspicture}\nonumber \\
&= \delta_{j0}\frac{d_k}{\cD} \ket{\Phi}_{e_1} \otimes \ket{0}_{e_2} \label{eq:tdrec}
\end{align}
where we have applied \lemref{t:punctureremoval}, and Eqs.~\eqnref{e:looprule} and~\eqnref{e:tadpolerule}.  Eq.~\eqnref{eq:bpproj} follows since $B_p$ is a projection.  

Now we can argue that $B_q^i Q_v B_p = Z_p^\dagger B_{q'}^i Z_p$, from which Eq.~\eqnref{eq:bptadpolecommutation} follows. On the left-hand side we know 
from~\eqnref{eq:bpproj} and~\eqnref{eq:tdrec}
\begin{align*}
Q_v B_p 
&= \proj{\Phi}_{e_1} \otimes \proj{0}_{e_2} \otimes \sum_i \proj{ii}_{e_3e_4} \nonumber \\
&= \cD B_p \, \ketbra{0}{\Phi}_{e_1} \otimes \proj{0}_{e_2} \otimes \Delta^\dagger \Delta
\end{align*}
where $\Delta = \sum_j \ket{j}_{e'} \bra{jj}_{e_3,e_4}$.  Similarly, we have
\begin{align*}
Z_p^\dagger B_{q'}^i Z_p
&= \cD B_p \, \Delta^\dagger B_{q'}^i \Delta \otimes \ketbra{0}{\Phi}_{e_1} \otimes \proj{0}_{e_2} \enspace . 
\end{align*}
Thus we need only verify that $B_p B_q^i \Delta^\dagger \Delta \otimes \ket{00}_{e_1e_2} = B_p \ket{00}_{e_1e_2} \otimes \Delta^\dagger B_{q'}^i \Delta \otimes \ket{00}_{e_1e_2}$.  Indeed, letting $\reduce_\cG$ (resp. $\reduce_{\cG'}$) mean reducing the smooth string net to $\cH_\cG$ (resp. $\cH_{\cG'}$), 
\begin{multline*}
B_p B_q^i(\id_{\cG\backslash\{e_3,e_4\}}\otimes\Delta^\dagger \Delta) \ket{00jj}_{e_1e_2e_3e_4} \\
\begin{split}
&=
B_p \;
\begin{pspicture}[shift=-.9](-.8,-1)(1.75,1)
\psset{unit=.7cm}
\psset{linewidth=.56pt} 
\psset{labelsep=2.5pt} 
\def\leglength{.5}
\def\leg #1#2{{\psline{-}(#1)([nodesep=\leglength,angle=#2]#1)}}
\SpecialCoor
\pnode(0,0){w}
\pnode([nodesep=.5,angle=0]w){p}
\pnode([nodesep=1.2,angle=0]p){q}
\pnode([nodesep=.7,angle=180]w){v}
\pnode([nodesep=1,angle=-90]v){v1}
\pnode([nodesep=1,angle=90]v){v2}
\leg{v2}{150}
\leg{v1}{-150}
\psline{-}(v1)([nodesep=1,angle=-30]v1)
\psline{-}(v2)([nodesep=1,angle=30]v2)
\ncline{->}{v1}{v} \Aput{$j$}
\ncline{->}{v}{v2} \Aput{$j$}
\ncline[linestyle=solid]{-}{v}{w} \Aput{$0$}
\psarc[linestyle=solid]{-}(p){.5}{-180}{180} \uput[120]([nodesep=.5,angle=120]p){$0$}
\rput(p){\puncture}
\rput([nodesep=5pt,angle=-30]p){$p$}
\rput(q){\puncture}
\rput([nodesep=5pt,angle=-30]q){$q$}
\psarc{->}(q){.5}{-210}{150}
\uput[120]([nodesep=.5,angle=120]q){$i$}
\end{pspicture} \\
&=
B_p \, \reduce_\cG \;
\begin{pspicture}[shift=-.9](-.8,-1)(1.75,1)
\psset{unit=.7cm}
\psset{linewidth=.56pt} 
\psset{labelsep=2.5pt} 
\def\leglength{.5}
\def\leg #1#2{{\psline{-}(#1)([nodesep=\leglength,angle=#2]#1)}}
\SpecialCoor
\pnode(0,0){w}
\pnode([nodesep=.5,angle=0]w){p}
\pnode([nodesep=1.2,angle=0]p){q}
\pnode([nodesep=.7,angle=180]w){v}
\pnode([nodesep=1,angle=-90]v){v1}
\pnode([nodesep=1,angle=90]v){v2}
\leg{v2}{150}
\leg{v1}{-150}
\psline{-}(v1)([nodesep=1,angle=-30]v1)
\psline{-}(v2)([nodesep=1,angle=30]v2)
\ncline{->}{v1}{v2} \Aput{$j$}
\rput(p){\puncture}
\rput([nodesep=5pt,angle=-30]p){$p$}
\rput(q){\puncture}
\rput([nodesep=5pt,angle=-30]q){$q$}
\pnode([nodesep=.75,angle=180]q){newcenter}
\psarc{->}(newcenter){1.25}{-210}{150}
\uput[-60]([nodesep=1.25,angle=120]newcenter){$i$}
\end{pspicture}\\
&=
B_p \ket{00}_{e_1e_2} \otimes \Delta^\dagger \reduce_{\cG'} \;
\begin{pspicture}[shift=-.9](-.8,-1)(1.75,1)
\psset{unit=.7cm}
\psset{linewidth=.56pt} 
\psset{labelsep=2.5pt} 
\def\leglength{.5}
\def\leg #1#2{{\psline{-}(#1)([nodesep=\leglength,angle=#2]#1)}}
\SpecialCoor
\pnode(0,0){w}
\pnode([nodesep=.5,angle=0]w){p}
\pnode([nodesep=1.2,angle=0]p){q}
\pnode([nodesep=.7,angle=180]w){v}
\pnode([nodesep=1,angle=-90]v){v1}
\pnode([nodesep=1,angle=90]v){v2}
\leg{v2}{150}
\leg{v1}{-150}
\psline{-}(v1)([nodesep=1,angle=-30]v1)
\psline{-}(v2)([nodesep=1,angle=30]v2)
\ncline{->}{v1}{v2} \Aput{$j$}
\pnode([nodesep=.75,angle=180]q){newcenter}
\rput(newcenter){\puncture}
\rput([nodesep=5pt,angle=-30]newcenter){$q'$}
\psarc{->}(newcenter){1.25}{-210}{150}
\uput[-60]([nodesep=1.25,angle=120]newcenter){$i$}
\end{pspicture} \\
&= 
B_p \left( \ket{00}_{e_1e_2} \otimes \Delta^\dagger B_{q'}^i \Delta \ket{jj}_{e_3e_4} \right)
\end{split}\end{multline*}
\noindent
where the first and last equalities are by definition of $B_q^i$ and $B_{q'}^i$, the second equality is by \lemref{t:tadpoleremoval}, and the third equality is because the exact same sequence of steps can be used to reduce the pictured smooth string net to $\cH_\cG$ as can be used to reduce it to $\cH_{\cG'}$.  
Eq.~\eqnref{eq:lemma3c} now follows immediately.  
\end{proof}


\end{document}